\documentclass[aps,preprint]{revtex4}
\usepackage{mathrsfs}

\usepackage{graphicx}
\usepackage{multirow}
\usepackage{makecell}
\usepackage{booktabs}

\usepackage{dcolumn}
\usepackage{bm}
\usepackage{xcolor}
\usepackage[utf8]{inputenc}
\usepackage[T1]{fontenc}

\newcommand{\bs}{Bi$_2$Se$_3$}

\newcommand{\rev}[1]{\textcolor{black}{#1}}

\begin{document}

\title{Angle-resolved photoemission spectroscopy with an \textit{in situ} tunable magnetic field}

\author{Jianwei Huang$^{1,*}$, Ziqin Yue$^{1,2}$, Andrey Baydin$^{3,4}$, Hanyu Zhu$^{1,5}$, Hiroyuki Nojiri$^{6}$, Junichiro Kono$^{1,3,4,5}$, Yu He$^{7,*}$, Ming Yi$^{1,*}$}
\affiliation{
\\$^{1}$Department of Physics and Astronomy, Rice University, Houston, Texas 77005, USA
\\$^{2}$Applied Physics Graduate Program, Smalley-Curl Institute, Rice University, Houston, Texas 77005, USA
\\$^{3}$Department of Electrical and Computer Engineering, Rice University, Houston, Texas 77005, USA
\\$^{4}$Smalley–Curl Institute, Rice University, Houston, Texas 77005, USA
\\$^{5}$Department of Materials Science and NanoEngineering, Rice University, Houston, Texas 77005, USA
\\$^{6}$Institute for Materials Research, Tohoku University, Katahira 2-1-1, Sendai 980-8577, Japan. 
\\$^{7}$Department of Applied Physics, Yale University, New Haven, Connecticut 06511, USA
\\$^{*}$To whom correspondence should be addressed: jwhuang@rice.edu, yu.he@yale.edu and mingyi@rice.edu
}


\begin{abstract}
 Angle-resolved photoemission spectroscopy (ARPES) is a powerful tool for probing the momentum-resolved single-particle spectral function of materials. Historically, \textit{in situ} magnetic fields have been carefully avoided as they are detrimental to the control of photoelectron trajectory during the photoelectron detection process. However, magnetic field is an important experimental knob for both probing and tuning symmetry-breaking phases and electronic topology in quantum materials.
 In this paper, we introduce an easily implementable method for realizing an \textit{in situ} tunable magnetic field at the sample position in an ARPES experiment and analyze magnetic field induced artifacts in ARPES data. Specifically, we identified and quantified three distinct extrinsic effects of a magnetic field: \rev{Constant energy contour rotation, emission angle contraction}, and momentum broadening. We examined these effects in three prototypical quantum materials, i.e., a topological insulator (\bs), an iron-based superconductor (LiFeAs), and a cuprate superconductor \rev{(Pb-Bi$_2$Sr$_2$CuO$_{6+x}$)}, and demonstrate the feasibility of ARPES measurements in the presence of a controllable magnetic field. Our studies lay the foundation for the future development of the technique and interpretation of ARPES measurements of field-tunable quantum phases.
\end{abstract}

\maketitle

\newpage

\section{Introduction}
Angle-resolved photoemission spectroscopy (ARPES) is a powerful technique that directly probes the electronic structure of quantum materials in momentum space. ARPES investigations have provided deep insights into the physical properties of a wide range of material systems, including heavy fermions, cuprate superconductors, iron-based superconductors, semiconductors, two-dimensional materials, and a variety of topological materials~\cite{Damascelli2003, Hufner2003, Lu2012, Sobota2021b}. For example, ARPES has played an irreplaceable role in investigating the superconducting mechanism of the copper oxides and iron-based superconductors, as well as their unusual normal states and symmetry-breaking competing orders~\cite{Damascelli2003, Campuzano2004, Yi2017d}. It has also revealed the key experimental signatures of topology in the band structures of topological materials~\cite{Dil2019, Lv2019a}, thereby contributing to the establishment of a complete set of topological band theory. As an energy-momentum-resolved one-electron spectral probe, ARPES has greatly advanced the frontier of condensed matter physics and quantum materials.

Reversely, rapid progress in condensed matter physics has driven the expansion of capabilities for modern ARPES~\cite{Sobota2021b}, ranging from novel light sources to high efficiency detection modes~\cite{Saitoh2000, Borisenko2012, Koralek2007, Liu2008, Kiss2008, He2016, Zhou2018}. Various ARPES techniques have been applied to study different materials, including nano-ARPES with sub-micron-sized beam spot, developed for studying mesoscopic and device physics~\cite{Dudin2010, Bostwick2012, Avila2014}; time-resolved ARPES (tr-ARPES), developed for studying non-equilibrium electronic states; and spin-resolved ARPES~\cite{Ishida2014}, developed for studying electron spin textures~\cite{Hoesch2002, Dil2009}. Additionally, there have been significant developments in the \textit{in situ} control of sample environments, such as surface alkaline metal dosing, uniaxial strain tuning, and electrostatic gating~\cite{Ohta2006, Ricco2018, Pfau2019, Cai2020, Nguyen2019, Joucken2019}. However, the implementation of magnetic control in the DC limit has been a longstanding challenge in the ARPES sample environment, which is extremely sensitive to magnetic field deflection effects of low kinetic energy photoelectrons~\cite{Kaminski2016}.

In an ARPES experiment, a light beam impinges on the sample, ejecting photoelectrons. The trajectories of the photoelectrons are kept unperturbed on their way to the electron analyzer, which, through a set of electrostatic lenses, guides the photoelectrons to the detector that measures the momentum and energy of the photoelectrons~\cite{Martensson1994}. As the effect of stray magnetic fields on photoelectron trajectories is difficult to predict and control, all ultra-high vacuum chambers used in ARPES experiments are designed to shield out magnetic fields, as they are detrimental to the energy-momentum resolvability of contemporary photoelectron analyzers. Consequently, the \textit{in situ} application of a  magnetic field on the sample during an ARPES experiment has long been considered impractical. In this paper, we introduce an easy-to-implement method for producing an \textit{in situ} controllable magnetic field via a solenoid at the sample position in our in-house ARPES system equipped with a helium lamp light source (Fig.~\ref{fig:Fig1}a). By applying an electric current to the solenoid, we are able to realize a tunable magnetic field with strength up to $\sim$10 mT with correctable photoelectron deflection effects at the He I$\alpha$ excitation energy (21.2~eV), making ARPES measurements under \textit{in situ} magnetic fields feasible. This excitation source is by far the most widely accessible and adopted high resolution deep UV source that is compatible in a lab-setting. However, the magnetic field deflection effect is expected to be considerably more pronounced and thus more challenging to handle than synchrotron photoemission, which is often operating at higher photon energies.
In the following, we will describe our setup, present our analysis of the effects of magnetic fields on photoelectrons, and demonstrate our ARPES measurements on three well-known quantum material systems using the developed setup.

\section{Electron trajectory analysis}
\begin{figure}
\includegraphics[width=0.98\textwidth]{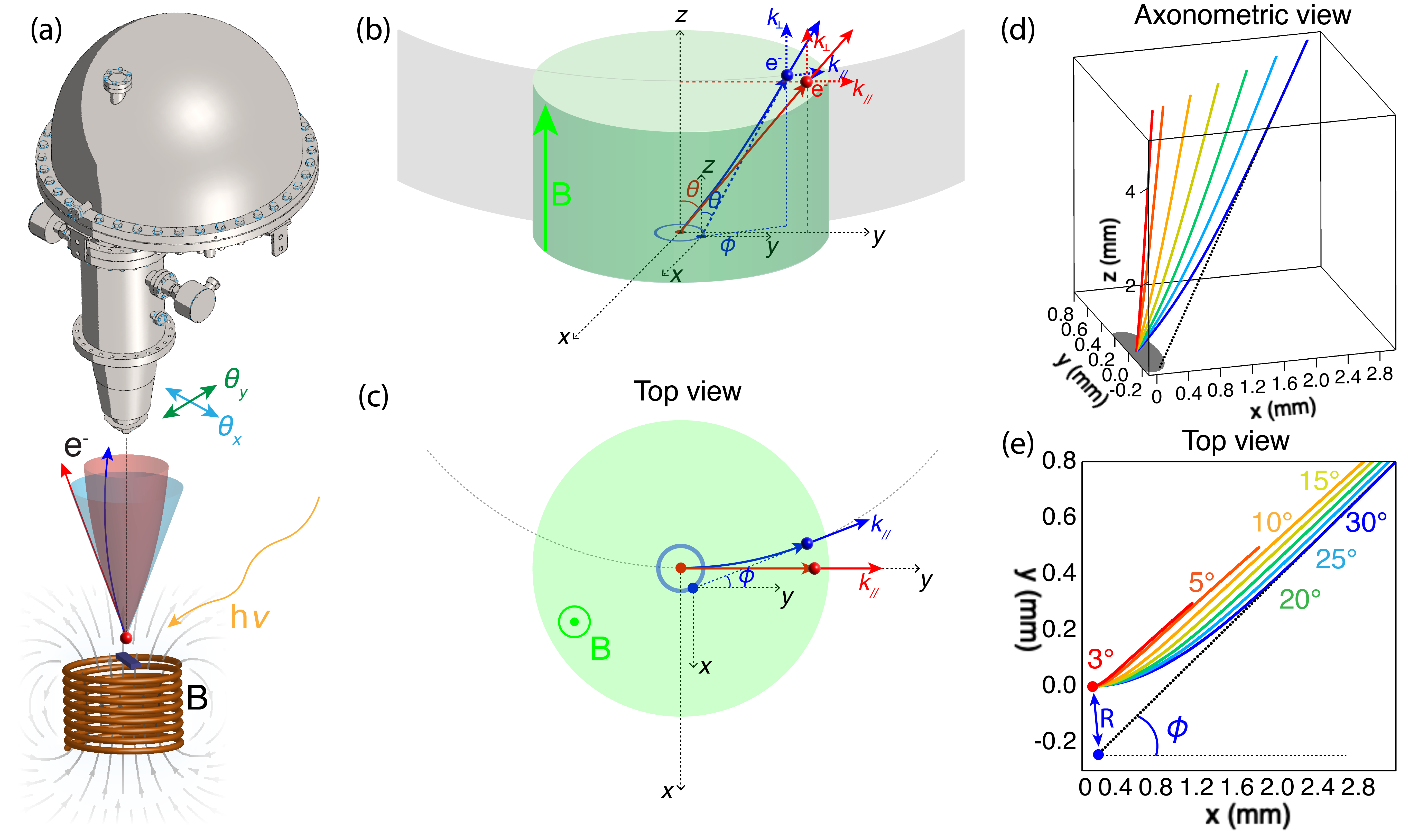}
\caption{\label{fig:Fig1}{\bf Photoelectron trajectory in a magnetic field and its effect on ARPES measurements.} (a) Schematic showing the experimental geometry of our ARPES measurements with a magnetic field. (b) Simplified model to analyze the effect of the magnetic field on ARPES experiments based on our experimental setup. The green shaded region shows an artificial constant magnetic field distribution restricted to a finite cylindrical region. The red arrow indicates the electron trajectory without the magnetic field. The blue arrow indicates the electron trajectory with the artificial magnetic field. $\theta$ and $\phi$ indicate the electron emission angle and azimuthal rotation. (c) Top view of (b). (d) Finite element analysis of the electron trajectory in a magnetic field produced by a cylindrical current. We use a cylindrical wall with 8 mm diameter and 4 mm height to simulate our solenoid. An electric current density of 5265 A/m is used to simulate a 0.25 A current applied on the 90-turn solenoid. The electrons were initially emitted within the \textit{xz} plane. The electron trajectories with different colors represent different emission angles, $\theta$, as indicated in (e). The grey spot on the basal plane ($z$ = 0) dictates the "virtual" spot. (e) Top view of (d) showing near-uniform azimuthal rotation angle (20$^\circ$) for photoelectrons from different emission angles. The enlarged virtual beamspot has a radius of R$\sim0.25$ mm.
}
\end{figure}

Considering that a magnetic field does not perturb the velocity of an electron along the field direction, the simplest type of magnetic field to consider is that of a field in the out-of-plane direction of the sample, easily implementable via a solenoid around the sample. For our experiments, we have built two such solenoid devices, a large one and a small one, with specific design parameters that will be introduced in a subsequent section. To gain an intuitive understanding of the effect of such a magnetic field on the energy-momentum map of a low-energy photoemission process, we first introduce a simplified analytical model. As the magnetic field drops off rapidly outside the solenoid along the solenoid axis, we assume the magnetic field ($B$) to be constant along the \textit{z}-axis that is confined within the green cylindrical region (Fig.~\ref{fig:Fig1}b). While this is not a realistic magnetic-field distribution, it captures the essence of the effects of a magnetic field on our ARPES measurements. Without loss of generality, we assume that an electron is emitted from the origin of our coordinate system within the \textit{yz}-plane at a finite angle $\theta$ with respect to the \textit{z}-axis and analyze its trajectory as it is affected by the magnetic field.

In the absence of a magnetic field, the electron follows a straight path indicated by the red arrow in Fig.~\ref{fig:Fig1}b. When a magnetic field is applied, the electron bends away from its original trajectory (the blue arrow in Fig.~\ref{fig:Fig1}b). This can be understood from the consideration of the Lorentz force, which causes the electron to bend within the plane perpendicular to the magnetic field (velocity component $v_{xy}$), but has no effect on the electron motion in the direction along the magnetic field (velocity component $v_z$). 
As a result, the electron travels in a helical trajectory confined within the grey wall (only a portion is shown in Fig.~\ref{fig:Fig1}b) until it exits the magnetic field region, after which it resumes a straight trajectory. From the top view, the trajectory traces out an arc (indicated by the grey dashed line in Fig.~\ref{fig:Fig1}c) followed by a tangential line after the electron leaves the magnetic field (blue arrow in Fig.~\ref{fig:Fig1}c). By the time the electron reaches the analyzer, it is perceived by the analyzer to have traveled along a "virtual" trajectory formed by a reverse linear extension of the trajectory outside the magnetic field region (blue dashed line in Fig.~\ref{fig:Fig1}b). From this "virtual" trajectory, we can see that the emission angle $\theta$ remains unchanged due to the unaffected magnitude of $v_{xy}$ and $v_z$ by the magnetic field. However, an in-plane rotation $\phi$ within the $xy$-plane is acquired, which is manifested in a \rev{constant energy contour (CEC)} rotation in ARPES measurements (Fig.~\ref{fig:Fig1}b and c). In addition, the electron is viewed by the analyzer as emitted from a different spot, which translates to an effectively enlarged spot size after considering all the electrons emitted with different emission angles (indicated by the blue circle in Fig.~\ref{fig:Fig1}b). Such an effectively enlarged spot size results in an overall momentum broadening effect due to the principles of the electrostatic focusing lens within the electron analyzer.

Next, we address the question of whether the \rev{CEC} rotation has radial shearing, i.e., whether the azimuthal rotation angle $\phi$ is strongly dependent on the emission angle $\theta$. We first perform an analysis based on our simplified model (Fig.~\ref{fig:Fig1}b). Since the electron's rotation within the $xy$-plane has an angular frequency ($\omega$) caused by the Lorentz force,
\begin{equation}
    \omega = \frac{eB}{m_e},
\end{equation}
which is independent of the electron velocity but only depends on the magnetic field $B$. The total rotation angle $\phi$ would be the amount by which the electron rotates before it exits the magnetic field region. If we assume that the magnetic field vanishes abruptly at $z = L$, then
\begin{equation}
    \phi = \frac{eB}{m_e}\frac{L}{v\cdot \cos(\theta)}.
\end{equation}
This seems to suggest that the \rev{CEC} rotation is $\theta$ dependent. However, since the acceptance angle of a standard ARPES analyzer is $\theta$ = $\pm$15$^\circ$, the largest rotation angle difference is only 3.5\%. For a larger emission angle of $\pm$30$^\circ$ used in our experimental examples, the largest rotation angle difference between different emission angles is 15\%. In practice, the angle difference could be smaller as the magnetic field generated by a solenoid weakens away from its center.

Now that we have gained intuition for the main effects of the magnetic field on the photoelectron trajectory -- a global azimuthal rotation and an enlarged virtual emissions spot, we use finite element analysis to quantitatively model and simulate the photoelectron trajectories in the magnetic field given the geometry of our physical setup. To generate a realistic magnetic field mimicking our test device, we used a cylindrical current based on the dimensions of the solenoid used in our ARPES system with an electric current of 0.25 A. Our simulated photoelectrons are emitted from the center of the top surface of the cylindrical current with different emission angles $\theta$. The electron kinetic energy used is 16.9 eV, corresponding to the largest kinetic energy of the photoelectrons excited by the He I$\alpha$ line in our system. The results, shown in Fig.~\ref{fig:Fig1}d and e, reveal that the trajectories of photoelectrons with different emission angles eventually become straight beyond $\sim$3 mm away from the center of the solenoid. Strikingly, the straight parts of the electron trajectories are almost parallel among a wide range of emission angles (up to 30$^\circ$ from normal emission), suggesting a uniformly rotated \rev{CEC} in the momentum maps of photoelectrons (Fig.~\ref{fig:Fig1}e). Our simulation suggests that the \rev{CEC} rotation caused by the magnetic field in our experimental setup can be treated as rigid.

\begin{figure}
\includegraphics[width=0.98\textwidth]{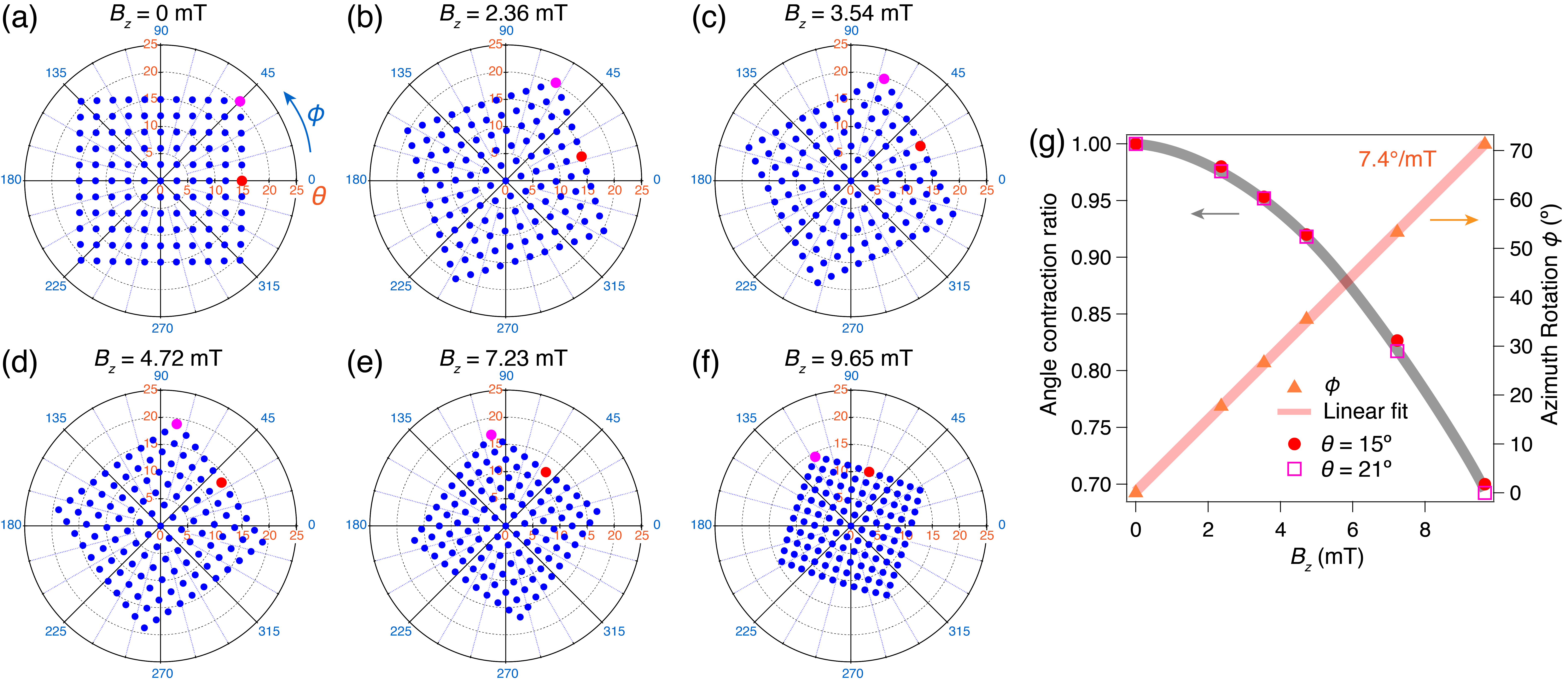}
\caption{\label{fig:Fig2}{\rev{\bf Simulation of photoelectron emission angle distribution maps in different magnetic fields.}} \rev{(a) A photoelectron emission angle distribution map plotted on a polar graph, where the radial direction is the emission angle ($\theta$) and the azimuthal angle is $\phi$. The points are evenly distributed in steps of 3$^\circ$ in the horizontal ($\theta_x$) and vertical ($\theta_y$) directions. (b-f) Simulations of photoelectron emission angle distribution maps of (a) after considering the corresponding magnetic fields (indicated in each panel) at the sample position generated by a cylindrical current. (g) The extracted emission angle ($\theta$) contraction and azimuth angle ($\phi$) rotation as a function of the magnetic field at the sample position from (a-f). The red solid circles and magenta empty squares are obtained from the corresponding red and magenta grid points as indicated in (a-f). The solid orange triangles indicate the photoelectron azimuthal rotation angle caused by the magnetic field and a linear fit gives 7.4$^\circ$/mT.}
}
\end{figure}

\rev{To directly visualize the aforementioned effects of the magnetic field on photoelectrons in an ARPES measurement directly, we simulated the photoelectron emission angle distribution maps under different magnetic field strengths generated by the same setup used above (Fig.~\ref{fig:Fig2}). We started with an evenly spaced photoelectron emission angle distribution map of 30$^\circ$ $\times$ 30$^\circ$ plotted on a polar graph (Fig.~\ref{fig:Fig2}a). The in-plane azimuthal angle is defined as $\phi$ and the radial emission angle is defined as $\theta$. The magnetic field strength is controlled by varying the electric current in the coil. The simulation demonstrates a clear rotation of the photoelectron map (Fig.~\ref{fig:Fig2}b-f), consistent with the previously demonstrated CEC rotation. Furthermore, the rotation angle is linearly proportional to the magnetic field strength as expected (Fig.~\ref{fig:Fig2}g). Although this simulation of the photoelectron emission angle map does not directly indicate an enlarged virtual emissions spot, we uncover another effect induced by the magnetic field — emission angle contraction. This effect is manifested by the gradual contraction of the photoelectron emission angle map (Fig.~\ref{fig:Fig2}a-f). We extracted the emission angle contraction ratios from two different emission angles and found them to follow the same trend as a function of the magnetic field (indicated by the grey solid line in Fig.~\ref{fig:Fig2}g). This suggests that the emission angle contraction is nearly uniform across various emission angles. The phenomenon of photoelectron emission angle contraction can be understood as a focusing effect of charged particle beams in a solenoid magnetic field~\cite{Kumar2009}.}

From the above analysis, we can summarize the extrinsic effects that exist in a typical He-I$\alpha$ ARPES experiment with an \textit{in situ} magnetic field produced by a solenoid: momentum-independent \rev{CEC rotation, momentum broadening and emission angle contraction. While we note that the field effects on electron trajectories and the electron lens system of the analyzer should strictly be treated as one imaging system, our analyses provided above is a way to derive intuitive understanding of the extrinsic effects of the magnetic field on the photoelectrons. Future analyses treating the complete photoelectron detection system including the lens system should be carried out for a more rigorous treatment of the effects.} Below we will address the identified three effects of magnetic field on ARPES experiments on real quantum materials in subsequent sections.

\section{Solenoid device design and characterization}
\begin{figure}
\includegraphics[width=0.8\textwidth]{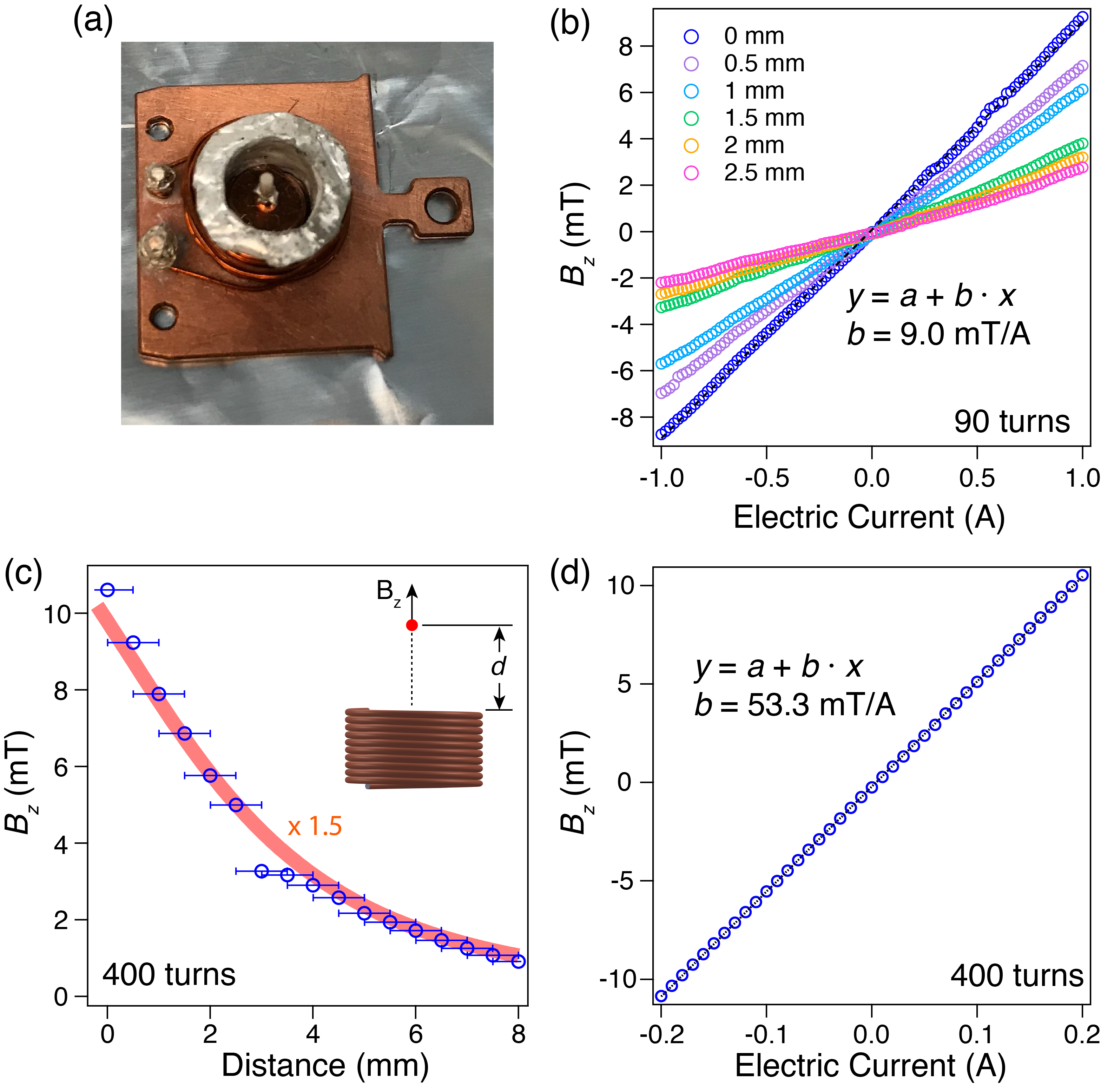}
\caption{\label{fig:Fig3}{\bf Characterization of the solenoid on the sample holder.} (a) A picture of our solenoid on a standard Omicron sample holder used in our home-lab system. (b) Characterization of the magnetic field ($B_z$) distribution by a 90-turn solenoid with varying electric current. The magnetic field was measured by a Hall bar. Labels with different colors indicate the distance from the top of the solenoid as indicated in the inset of (c). (c) $B_z$ as a function of the distance from a 400-turn solenoid on a sample holder. The red shaded curve is a finite element analysis based on our solenoid dimension applied with an electric current of 0.2 A. (d) Characterization of the magnetic field ($B_z$) distribution by a 400-turns solenoid with varying electric current. A linear fit of (b) and (d) could give the magnetic field ($B_z$) based on the electric current applied.
}
\end{figure}

To implement a tunable out-of-plane magnetic field, our sample device design consists of a single solenoid made of circular winding of a kapton-insulated copper wire (Kurt J. Lesker), the ends of which are connected to pins on a modified flag-style Omicron sample holder that can be inserted into the sample manipulator with electrical contacts. A single crystalline sample can be mounted on a standard copper holder that is screwed into the center of the solenoid, with a sample height near the top of the solenoid. A tunable magnetic field can be applied at the sample position via the control of the current in the solenoid (Fig.~\ref{fig:Fig1}a). For testing purposes, we made two prototype solenoid devices. A smaller solenoid consists of 90 turns using a wire with a diameter of 0.25 mm, resulting in a solenoid with an inner diameter of 6 mm, an outer diameter of 11 mm, and a height of 4 mm. A larger solenoid consists of 400 turns using a wire diameter of 0.14 mm, resulting in a solenoid with an inner diameter of 6 mm, an outer diameter of 13 mm, and a height of 6 mm. Figure~\ref{fig:Fig3}a shows a picture of the small solenoid mounted on a flag-style Omicron sample holder. 

We show the characteristics of the magnetic field generated at the sample position and its spatial profile along the axial direction of each coil (Fig.~\ref{fig:Fig3}b-d) measured using a Keithley 2400 source meter and a Hall sensor (HE244T from Asensor Technology AB). As expected, the magnetic field along the \textit{z}-axis ($B_z$) measured at the sample position is a linear function of the applied electric current ($I$). For the small solenoid (Fig.~\ref{fig:Fig3}b), a linear fit gives a slope of 9 mT/A, which allows us to calculate the \textit{in situ} magnetic field at the sample position from the electric current applied during the measurement. 
The same linear behavior of the magnetic field against the electric current is observed away from the solenoid along the $z$-axis (Fig.~\ref{fig:Fig3}b), with a decreasing slope. Similar characterization of the large solenoid gives a $B_z/I$ ratio up to 53.3 mT/A, allowing us to obtain a maximum magnetic field of 13 mT at the sample position produced by a 0.25 A current (Fig.~\ref{fig:Fig3}c and d). We plot $B_z$ as a function of the distance away from the solenoid in Fig.~\ref{fig:Fig3}c to compare with the finite element analysis (red line). The measured results show great consistency with the simulation predictions.

\section{Demonstration of ARPES measurements on quantum materials in a magnetic field}
Finally, we present three examples of ARPES measurements with \textit{in situ} magnetic fields. We selected three compounds that represent different areas of interest in condensed matter physics: \bs~for topological materials, LiFeAs for iron-based superconductors, and \rev{Pb-Bi$_2$Sr$_2$CuO$_{6+x}$ (Pb-Bi2201)} for cuprate superconductors. For each example, we will demonstrate the impact of the three extrinsic effects induced by the \textit{in situ} magnetic field on the measured electronic structure by ARPES. These results provide critical information for us to distinguish the intrinsic responses of the sample from the extrinsic effects of the apparatus caused by the magnetic field in ARPES experiments. All ARPES data presented here were taken with an in-house ARPES system with a Scienta DA30 electron analyzer and He-I$\alpha$ light source (Fermion Instruments, BL1100S), in ultra-high vacuum chamber with a base pressure of 5 $\times$ 10$^{-11}$ Torr. The maps presented were all taken with the sample rotated with respect to the analyzer unless otherwise noted.

\subsection{\bs}
\begin{figure}
\includegraphics[width=0.95\textwidth]{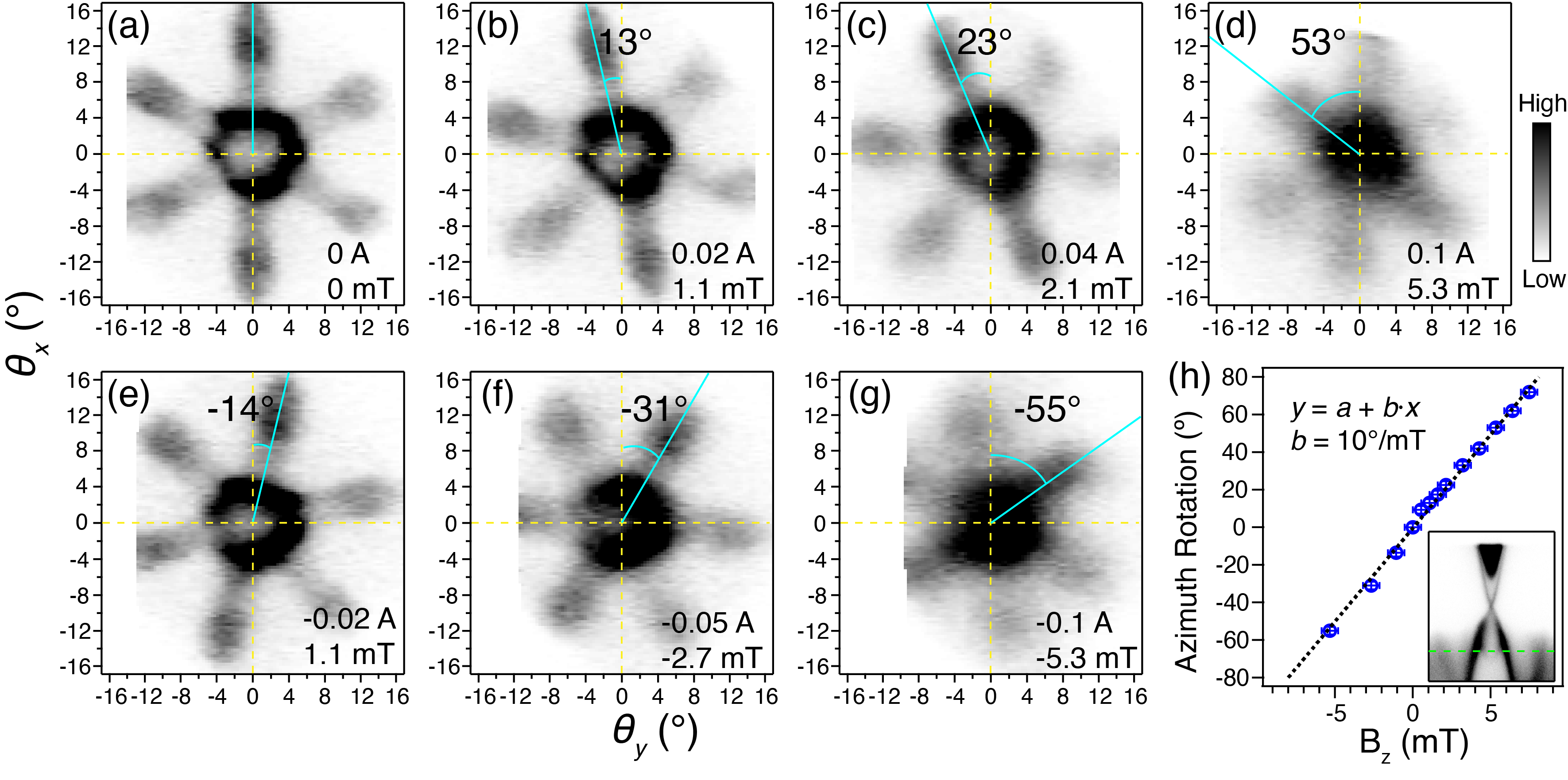}
\caption{\label{fig:Fig4}{\bf ARPES measurement of \bs~in a magnetic field.} (a) Constant energy contour (CEC) at the energy position indicated by the green dashed line in (h) of \bs~measured using a helium lamp (21.2 eV) without a magnetic field. (b-d) Same as (a) but with a magnetic field with different values by varying the applied electric current. The solid cyan line is an indication of in-plane rotation of the measured CECs. (e-g) Same as (b-d) but reversing the electric current. The corresponding in-plane rotation is also reversed. The measured spectra become broadened, along with contraction, as observed by ARPES under the influence of a magnetic field. (h) In-plane rotation of the CECs measured by ARPES as a function of applied magnetic field at the sample position. All maps taken with DA30 deflector mode.
}
\end{figure}

\bs~is a prototypical topological insulator~\cite{Zhang2009a}. Its electronic structure is well understood, making it a suitable sample to examine and characterize the extrinsic effects caused by the magnetic field in ARPES measurements. The band dispersions along the M-$\Gamma$-M is shown in the inset of Fig.~\ref{fig:Fig4}h, where the Dirac topological surface state can be seen within a bulk gap. To clearly demonstrate the effect of the magnetic field, we show the CEC at 0.7 eV binding energy, measured under a magnetic field produced by the large solenoid (Fig.~\ref{fig:Fig4}). The applied electric current and the corresponding magnetic field are shown for each CEC. For a current of 0.02 A (1.1 mT), a CEC rotation of 13$^\circ$ was observed (Fig.~\ref{fig:Fig4}b). With increasing current, the rotation angle increases, while reversing the electric current also reverses the magnetic field and therefore the CEC rotation angle (Fig.~\ref{fig:Fig4}e-g). The extracted CEC rotation angle scales linearly with the magnetic field (Fig.~\ref{fig:Fig4}h). A linear fit of the rotation angle versus field gives a ratio of 10 $^\circ$/mT for the large solenoid used here. It should be clarified that this value is not universal for all ARPES experiments with \textit{in situ} magnetic fields, as it varies between solenoids with different dimensions due to the different magnetic field distributions. However, such a simple geometry can be modeled and reproduced to a high precision with the finite element analysis shown above.

In addition to the CEC rotation, we also observe momentum broadening and emission cone contraction effects in this experiment. As shown in Fig.~\ref{fig:Fig4}a-g, the spectra of the central hexagon become increasingly broadened as the magnetic field is increased, as do the six lobes. At the same time, the six lobes move closer to the center as the magnetic field is increased. These observations confirm that the magnetic field causes momentum broadening and \rev{photoelectron emission angle contraction} effects in ARPES measurements, which we will quantify in the following sections.

\subsection{LiFeAs}
\begin{figure}
\includegraphics[width=0.8\textwidth]{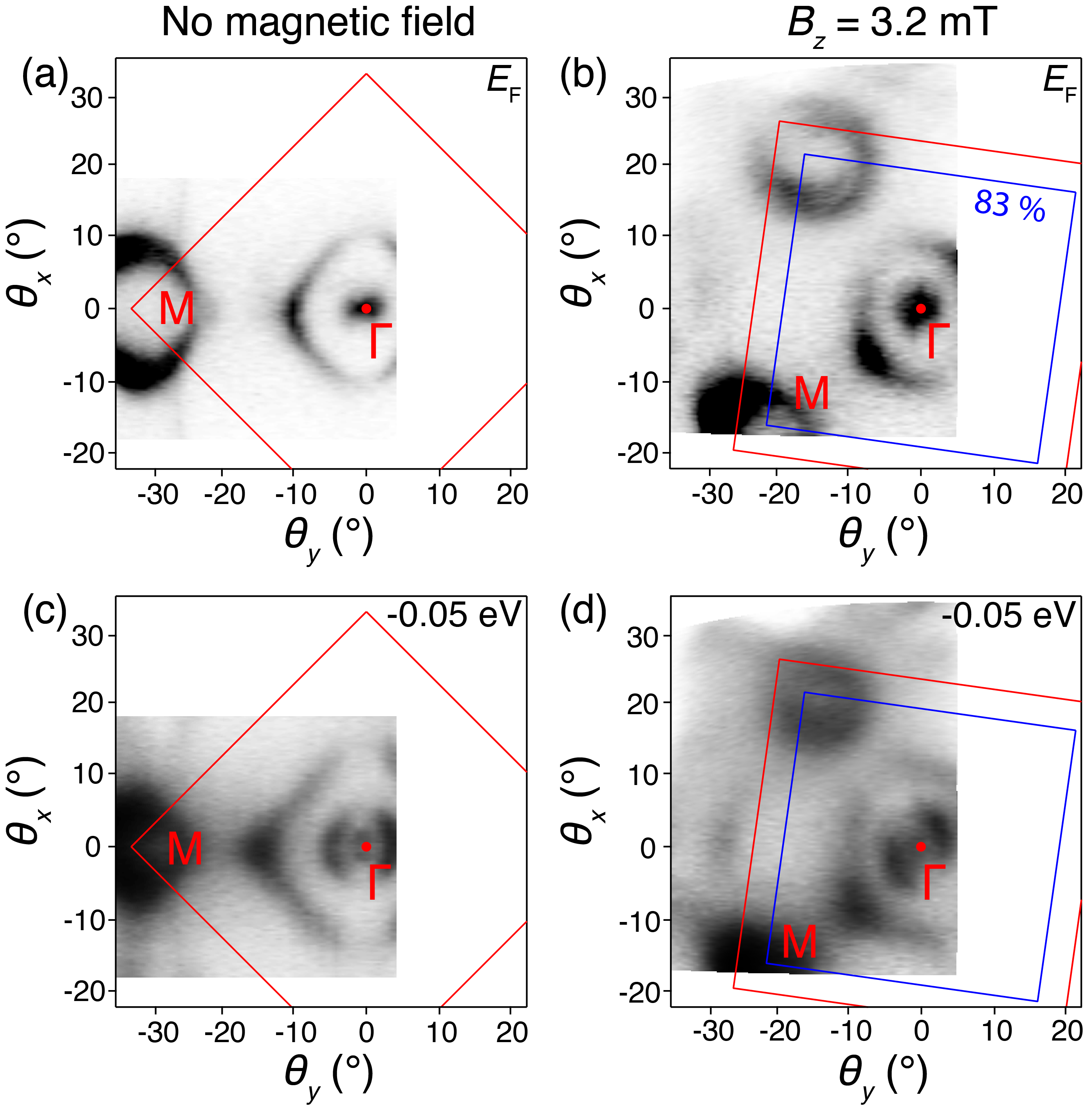}
\caption{\label{fig:Fig5}{\bf ARPES measurement of LiFeAs in a magnetic field.} (a) Fermi surface of LiFeAs measured by ARPES without applying magnetic field. (b) Fermi surface of LiFeAs measured by ARPES with a 3.2 mT magnetic field at the sample position. An overall Fermi surface rotation of 37$^\circ$ and emission angle contraction of 83$\%$ are observed. (c) CEC at -0.05 eV of LiFeAs without a magnetic field. (d) CEC at -0.05 eV measured by ARPES under the same condition as (b). Identical rotation and angle contraction effects are observed between Fermi surface and CEC at -0.05 eV.
}
\end{figure}

LiFeAs is a well-known iron-based superconductor that has a Fermi surface consisting of hole pockets around the Brillouin zone (BZ) center ($\Gamma$) and electron pockets around the BZ corner (M)~\cite{Umezawa2012}. This particular Fermi surface topology is suitable for characterizing the \rev{emission angle contraction} effect in ARPES measurements with an \textit{in situ} magnetic field. The data were collected with the large solenoid device. We first present the Fermi surface of LiFeAs measured without a magnetic field (Fig.~\ref{fig:Fig5}a). Small circular hole and electron pockets are clearly identified, with the corresponding momentum positions that match well with the BZ size. After applying an electric current of 0.06 A to generate a 3.2 mT magnetic field at the sample position, the three extrinsic effects are clearly observed from the Fermi surface mapping (Fig.~\ref{fig:Fig5}b): Fermi surface rotation; broadening of both the hole and electron Fermi pockets; and \rev{contraction} of the Fermi surfaces. From the location of the Fermi pockets, the \rev{angle contraction} effect can be quantitatively determined. In the Fermi surface mapping taken with a 3.2 mT magnetic field at the sample position, the Fermi surface rotates by 37$^\circ$ and \rev{contracts} to 83\% of its original BZ size without a magnetic field (indicated by the blue square in Fig.~\ref{fig:Fig5}b). We further examine these effects at a slightly higher binding energy (Fig.~\ref{fig:Fig5}c, d). Both the \rev{CEC} rotation and the \rev{emission angle contraction} effect do not exhibit an obvious dependence on the binding energy in the energy range relevant to low energy dispersions in LiFeAs.

\subsection{Pb-Bi$_2$Sr$_2$CuO$_{6+x}$}
The third example is \rev{Pb-Bi$_2$Sr$_2$CuO$_{6+x}$ (Bi2201)}, which is a prototypical cuprate superconductor. Its electronic structure exhibits a dichotomy between nodal and antinodal regions. One well-established feature is that the band dispersion around the nodal region is much steeper than that in the antinodal region~\cite{Kaminski2005}. In this section, we use Bi2201 to study the extrinsic effects caused by a magnetic field on spectral lineshape in ARPES. This measurement was carried out using the small solenoid with an electric current of 0.23 A, which generated a 2.1 mT magnetic filed at the sample position.

\begin{figure}
\includegraphics[width=0.95\textwidth]{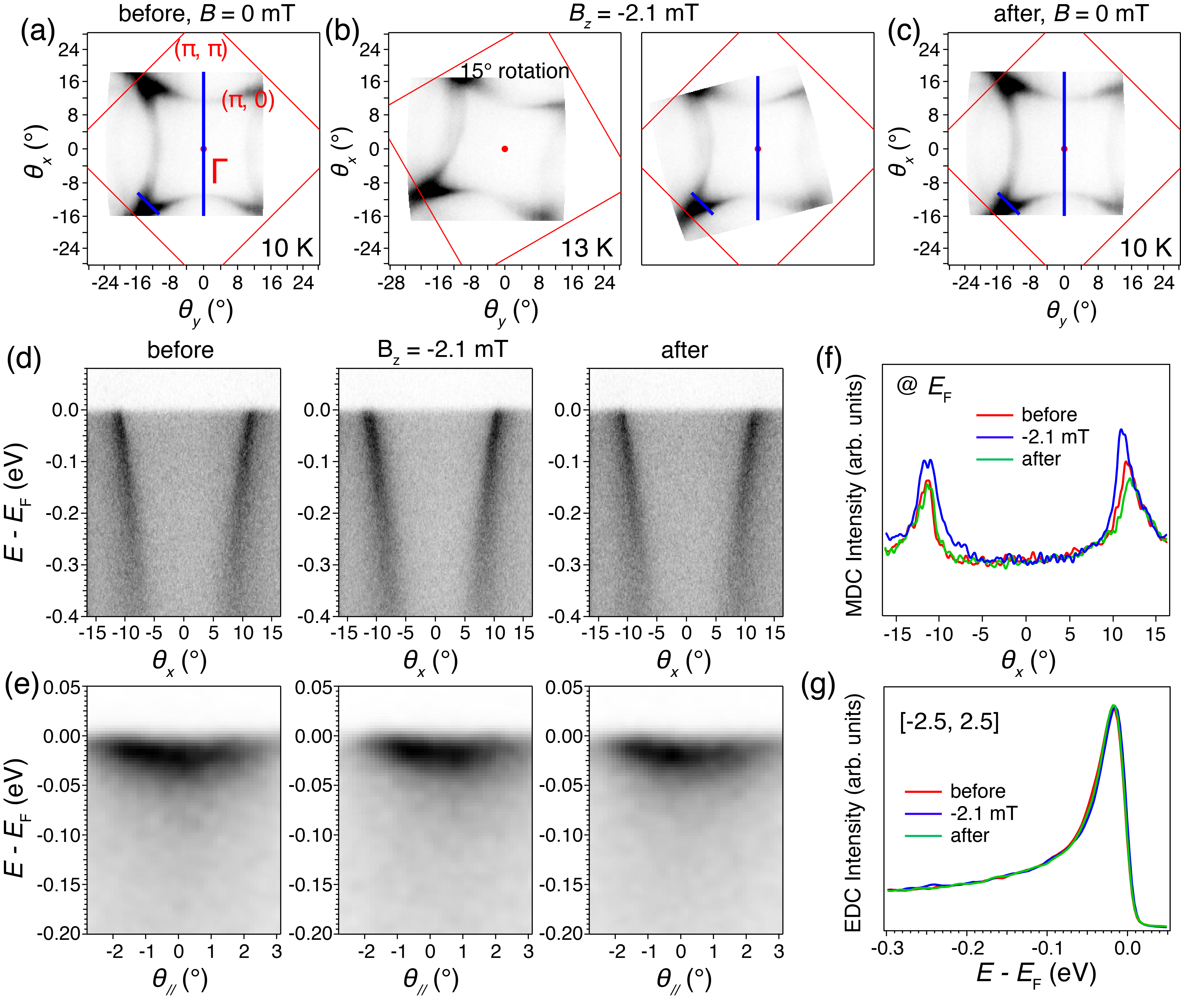}
\caption{\label{fig:Fig6}{\bf ARPES measurement of Pb-Bi$_2$Sr$_2$CuO$_{6+x}$ (Pb-Bi2201) in a magnetic field.} (a) Fermi surface mapping of Pb-Bi2201 measured by ARPES before applying a magnetic field. (b) Left: Same as (a) but applying a 2.1 mT magnetic field at the sample position. Fermi surface rotation of 15$^\circ$ is revealed. Right: Re-align the Fermi surface mapping by compensating for the field-induced Fermi surface rotation. (c) Fermi surface mapping after switching off the magnetic field. (d) Spectral images of the nodal cut indicated in (a-c) with and without magnetic field. (e) Spectral images of the anti-nodal cut indicated in (a-c) with and without a magnetic field. (f) Momentum distribution curves (MDCs) at $E_F$ of (d). (g) Integrated energy distribution curves (EDCs) of (e).
}
\end{figure}

The Fermi surface mappings without and with a magnetic field are shown in Figs.~\ref{fig:Fig6}a and b, respectively, integrated within a 20 meV energy window around the Fermi level. A clear Fermi surface rotation of 15$^\circ$ is observed caused by the magnetic field (Fig.~\ref{fig:Fig6}b). Subsequently, we switched off the magnetic field and repeated the measurement without a field to check the results (Fig.~\ref{fig:Fig6}c). We noticed a slight increase in the sample temperature due to the Joule heating of the solenoid caused by the electric current, which can be compensated by opening up the helium flow to increase the cooling power. \rev{We can achieve a temperature lower than 12 K at the sample stage while applying a current of 0.2 A on the large solenoid.} After correctly realigning the Fermi surface map with the magnetic field, we took the corresponding cuts along the nodal direction and around the antinodal region (Fig.~\ref{fig:Fig6}d, e). The spectral images along the nodal direction show a steep band dispersion, with a moderate momentum broadening effect measured with a magnetic field. The momentum broadening effect can be viewed more clearly from the momentum distribution curves (MDCs) around the Fermi level (Fig.~\ref{fig:Fig6}f). MDC measured with the magnetic field (blue line) has a larger peak width than that without a magnetic field (red and green lines). The momentum broadening effect not only broadens the line-width along the momentum direction, but also affects the lineshape of spectra along the energy direction, especially when the electronic structure has a steep band dispersion, which we will discuss in the next section. However, it has a negligible effect on the energy distribution curve (EDC) when the band dispersion is flat. As shown in the cuts around the antinodal region of Bi2201 (Fig.~\ref{fig:Fig6}e), the spectral images exhibit a rather flat band dispersion and the corresponding EDCs (Fig.~\ref{fig:Fig6}g) are almost identical with and without a magnetic field.

\section{Magnetic field-Induced extrinsic effects on ARPES spectral analysis}
Through the above analysis and measurements, we demonstrated that three extrinsic effects are introduced to ARPES measurements by a magnetic field, i.e., CEC rotation, momentum broadening, and \rev{emission angle contraction}. We examined these effects on the spectra of three prototypical examples measured by ARPES. The key findings are:

$\bullet$ {\bf CEC rotation:} CEC rotation is linearly proportional to the magnetic field strength at the sample position. Within the range of interest in a typical ARPES experiment, CEC rotation appears to be nearly uniform as a function of both momentum and energy.

$\bullet$ {\bf \rev{Emission angle contraction}:} Within the range of interest in a typical ARPES experiment, \rev{emission angle contraction} appears to be nearly uniform as a function of both momentum and energy.

$\bullet$ {\bf Momentum broadening:} Momentum broadening causes an increase in the MDC line-width of ARPES spectra, and does so in two-dimensional (2D) momentum space.

Compensating for the CEC rotation and \rev{emission angle contraction} effects is relatively straightforward, as they are practically uniform as a function of both momentum and energy within the typical energy-momentum range probed by a lab-based He-I$\alpha$ excitation source. The momentum broadening effect convolves information in the unperturbed ARPES spectra, and is more challenging to remove because it originates from an enlarged virtual emission spot that gradually degrades the focusing ability of the preset electrostatic lens table. Advanced numerical deconvolution algorithms, an accurate experimental measure of the point spread function from a point emission source, or a redesigned lens table that systematically includes the coil as the first lens element, may help mitigate this artifact. At the current level, we describe our observation of the momentum broadening effect that could potentially cause non-trivial artifacts in ARPES data.

Since a magnetic field does not alter the kinetic energy of the photoelectrons, na\"ively one might assume that the spectra along the energy direction measured by ARPES are not affected by the magnetic field. However, the energy and momentum line profiles are intimately linked in ARPES. We caution that momentum broadening alone could still affect both MDC and EDC lineshapes, potentially leading to incorrect conclusions from experimental data measured with magnetic fields. Here, we use \bs~as an example to demonstrate the situation where the momentum broadening effect could lead to a false conclusion on the opening of a band gap.

\begin{figure}
\includegraphics[width=0.95\textwidth]{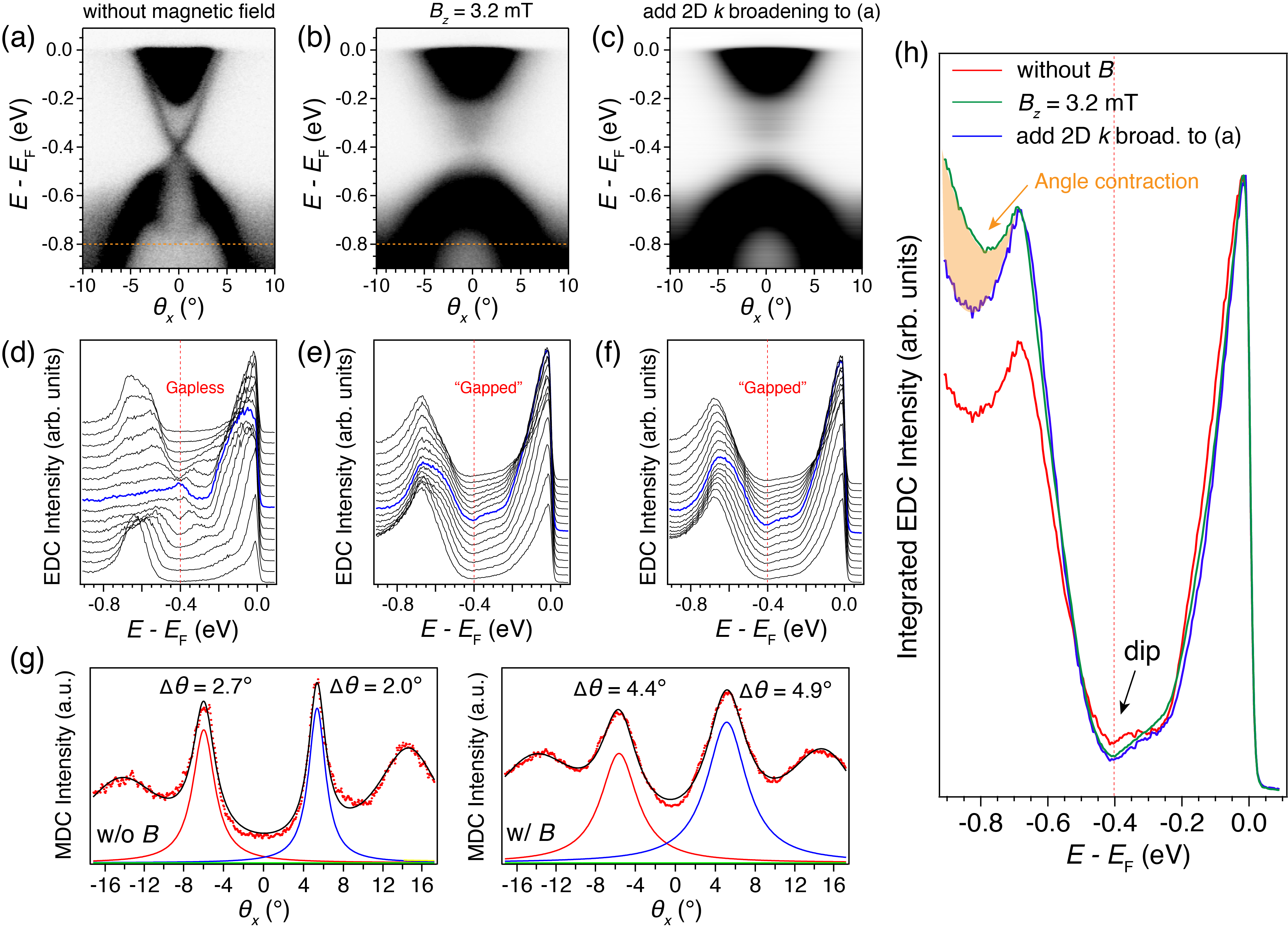}
\caption{\label{fig:Fig7}{\bf Effect of momentum-broadening on the Dirac point measured on \bs~in a magnetic field.} (a) Spectral image along M-$\Gamma$-M of \bs~measured without a magnetic field. (b) Spectral image along M-$\Gamma$-M of \bs~measured after turning on a 3.2 mT magnetic field at the sample position. (c) Spectral image along M-$\Gamma$-M obtained by applying a 2D momentum broadening to a 3D cube constructed from revolving the spectral in (a), assuming an isotropic spectral distribution. The amount of momentum broadening is obtained by comparing the line-width of MDCs of (a) and (b). (d-f) EDC stacks of (a-c). The blue EDCs correspond to those at $\Gamma$. The red dashed lines indicate the energy position of the Dirac point. The EDC stacks from spectral image of (a) show a gapless Dirac surface state while both (b) and (c) show a "gapped" surface state. (g) MDC at -0.8 eV of (a) and (b) for estimating the momentum broadening. The corresponding peak widths are fitted with Lorentzian functions. (h) Integrated EDCs of (a-c). All these EDCs are normalized at $E_F$. There is an extra loss of spectral intensity at the Dirac point in a single cut measured by ARPES due to the 2D momentum broadening. The measured intensity at high binding energy (-0.8 eV) with the magnetic field is larger because of the additional contribution from the electronic states at large momenta due to the emission angle contraction effect.
}
\end{figure}

Spectral images along the M-$\Gamma$-M direction of \bs~measured without and with a magnetic field are shown in Fig.~\ref{fig:Fig7}a and b, respectively. From the spectral image and corresponding EDC stacks obtained without a magnetic field, clear gapless Dirac cone surface states are observed within the bulk band gap (Fig.~\ref{fig:Fig7}a and d). In the spectra taken under a field of 3.2 mT, the spectra become significantly broadened and a "gap" feature appears from both the spectral image (Fig.~\ref{fig:Fig7}b) and as a dip in the corresponding EDC at the $\Gamma$ point (Fig.~\ref{fig:Fig7}e). The critical issue is whether the "gap opening" is an intrinsic effect of the magnetic field on the surface states or an extrinsic effect of momentum broadening caused by the magnetic field. To examine this, we note that the momentum broadening associated with the applied magnetic field occurs along 2D while our spectrum is a one-dimensional slice of the 2D momentum space. In order to correctly mimic the broadening effect, we first generate a three-dimensional (3D) cube from the spectral image in Fig.~\ref{fig:Fig7}a taken without a magnetic field. The 3D cube is generated artificially by assuming a circular revolution of the cut, then convolved with a Gaussian function with a linewidth determined by the observed momentum broadening ($\Delta$$\theta$ = 4.4$^\circ$) obtained from the measured MDCs without and with the magnetic field (Fig.~\ref{fig:Fig7}g). We then extract a slice from the momentum-broadened cube (Fig.~\ref{fig:Fig7}c) and observe striking similarity with the spectrum taken under the field (Fig.~\ref{fig:Fig7}b). The corresponding EDCs (Fig.~\ref{fig:Fig7}f) also exhibit a "gap opening" effect. We note that the integrated EDC of a single cut from the data without a magnetic field but convolved with momentum broadening matches well with that from the data measured with a magnetic field (blue and green lines in Fig.~\ref{fig:Fig7}h, where the deviation at high binding energy is due to the \rev{angle contraction} effect). This simulation suggests that the apparent gap opening spectrum shown here at the Dirac point is likely caused by a momentum broadening effect, instead of an intrinsic behavior of the sample induced by the magnetic field. This effect can be intuitively understood as the 2D momentum broadening spreads the spectral intensity associated with a Dirac point in both momentum directions, hence reducing the intensity along a single slice. This exercise cautions against drawing conclusions of field-induced gap opening based on data taken on a single slice. Instead, comparison on 2D maps should be made to fully exclude the 2D momentum broadening effect when addressing a potential gap opening effect associated with a Dirac point.

\section{Summary and outlook}
To summarize, we have introduced a method to apply an \textit{in situ} tunable magnetic field at the sample position using a small solenoid for ARPES measurements. The device is simple to implement on standard flag-style sample holders. By analyzing the electron trajectories with a toy model as well as with realistic finite element analysis, we identified three main extrinsic effects caused by the out-of-plane magnetic field generated by the solenoid: CEC rotation, momentum broadening, and \rev{emission angle contraction}. We demonstrated and evaluated these effects on three prototypical quantum materials, including \bs, LiFeAs, and Bi2201. Finally, we cautioned against potential pitfalls when interpreting ARPES measurements under an  \textit{in situ} magnetic field.

Importantly, our work demonstrates the feasibility of using an \textit{in situ} tunable magnetic field in ARPES experiments. 
Currently, we are able to obtain reasonable ARPES spectra in magnetic fields up to 3.2 mT at the sample position in our system. \rev{While the modest field we can realize is currently too small to observe field-induced Zeeman splitting of non-magnetic Dirac system, the simple coil design opens up opportunities for novel types of measurements. For magnetic topological materials, for example, a magnetic field can induce spin rotation that can lead to the opening or closing of Dirac gaps or cause topological phase transitions~\cite{Kim2018, Nie2020}. The design also allows} \textit{in situ} field-cooling preparation of magnetic samples, or even be utilized to have \textit{in situ} control of magnetic substrates, resulting in a stronger but more confined magnetic field that can be used for ARPES measurements. 
With an understanding of the extrinsic effects due to a magnetic field, it is possible to extract the intrinsic electronic response of a material to the magnetic field. Our work motivates the possibility to design electron analyzer modes that consider the out-of-plane field towards conditions where the extrinsic effects could be mitigated. 
Given the demonstrations of the feasibility of incorporating an \textit{in situ} magnetic field in the ARPES sample environments, we are optimistic that one day, ARPES could directly contribute to the study of magnetic field tuning of quantum phases.

While preparing this manuscript, we became aware of a related work that has appeared on arXiv, which explores "magnetoARPES", a variant of ARPES that can be conducted in a magnetic field at synchrotron x-ray photon energies \rev{with a different type of sample device}~\cite{Ryu2023}. 


\section{Acknowledgments}
We acknowledge the funding support of the Gordon and Betty Moore Foundation’s EPiQS Initiative through grant no. GBMF9470, and the NSF MRI grant no. DMR-2019004. Y.H. acknowledge the funding support of NSF CAREER award no. DMR-2239171. A.B. and J.K. acknowledge support from the Robert A. Welch Foundation through grant number C-1509.

\section{Author contributions}
J.H., Y.H., and M.Y. conceived the idea. J.H. designed the device and performed the ARPES experiments under the guidance of A.B., H.Z., H.N., J.K., Y.H., and M.Y. J.H. conducted the analysis of the toy model for photoelectron trajectory and simulation. Y.H. wrote the code for finite element analysis. J.H. analyzed the data under the guidance of Y.H. and M.Y. J.H. and Z.Y. made the figures. J.H., Y.H., and M.Y. wrote the paper with contributions from all authors.

\section{Data Availability}
All data needed to evaluate the conclusions are present in the paper. Additional data are available from the corresponding authors on reasonable request.

\section{Competing interests}
The authors declare no competing interests.

\end{document}